\documentclass[prl,showpacs,aps,twocolumn]{revtex4}
\usepackage{graphicx}
\usepackage{epstopdf}
\usepackage{dcolumn}
\usepackage{bm}
\usepackage{color}
\usepackage{ulem}
\usepackage{amsfonts,amsthm}
\usepackage{amsmath}
\usepackage{amssymb}
\begin{document}
\newcommand{\bR}{\mbox{\boldmath $R$}}
\newcommand{\Ha}{\mathcal{H}}
\newcommand{\mh}{\mathsf{h}}
\newcommand{\mA}{\mathsf{A}}
\newcommand{\mB}{\mathsf{B}}
\newcommand{\mC}{\mathsf{C}}
\newcommand{\mS}{\mathsf{S}}
\newcommand{\mU}{\mathsf{U}}
\newcommand{\mX}{\mathsf{X}}
\newcommand{\sP}{\mathcal{P}}
\newcommand{\sL}{\mathcal{L}}
\newcommand{\sO}{\mathcal{O}}
\newcommand{\la}{\langle}
\newcommand{\ra}{\rangle}
\newcommand{\ga}{\alpha}
\newcommand{\gb}{\beta}
\newcommand{\gc}{\gamma}
\newcommand{\gs}{\sigma}
\newcommand{\vk}{{\bm{k}}}
\newcommand{\vq}{{\bm{q}}}
\newcommand{\vR}{{\bm{R}}}
\newcommand{\vQ}{{\bm{Q}}}
\newcommand{\vga}{{\bm{\alpha}}}
\newcommand{\vgc}{{\bm{\gamma}}}
\newcommand{\mb}[1]{\mathbf{#1}}
\def\vec#1{\boldsymbol #1}
\arraycolsep=0.0em
\newcommand{\Ns}{N_{\text{s}}}
%

\title{
Superconductivity Emerging from Excitonic Mott insulator \\
--- {Theory of} Alkaline Doped Fullerene
}

\author{
Takahiro Misawa$^1$ and Masatoshi Imada$^2$
}

\affiliation{$^1$Institute for Solid State Physics, University of Tokyo, 5-1-5 Kashiwa-noha, Kashiwa, Chiba 277-8581, Japan}
\affiliation{$^2$
Department of Applied Physics, University of Tokyo,
7-3-1 Hongo, Bunkyo-ku, Tokyo, 113-8656, Japan
}

\date{\today}

\begin{abstract}
{A three-orbital model {derived} from the two-dimensional 
projection of the {\it ab initio} Hamiltonian for alkaline 
doped fullerene A$_3$C$_{60}$ with A=Cs,Rb,K is studied by a variational Monte Carlo method. We correctly reproduce the experimental 
isotropic $s$-wave superconductivity 
around the {\it ab initio} parameters. 
{With narrowing the bandwidth, the transition to an insulator is also reproduced, where orbital symmetry {is found to be spontaneously broken with emergence of} an excitonic Mott insulator for two orbitals and an antiferromagnetic insulator {nearly} degenerate with a spin liquid for the third orbital. The superconductivity is {a consequence of exciton melting}.      
}}
\end{abstract}

\pacs{
74.20.Pq,
74.70.Wz,
75.10.Kt
}

\maketitle

{\it Introduction.}--High critical temperature (high-$T_{c}$) 
superconductivity induced by
strong electronic correlations is one of the central topics
in condensed matter physics and 
it has been intensively studied since its
discovery in a family of cuprate compounds~\cite{Bednorz,ImadaRMP,LeeRMP},
{where crucial role of strong electron correlations are prominent in various aspects.} 
Recent discovery of the high-$T_{c}$ superconductivity in the 
iron-based superconductors~\cite{Kamihara_LaFeAsO},
where the five iron 3$d$ orbitals 
contribute to the low-energy degrees of freedom,
renewed the interest of the multi-orbital effects 
on the high-$T_{c}$ superconductivity.
Multi-orbital physics such as Hund's physics as well as
the orbital differentiation
was proposed to play essential roles for understanding 
the normal state properties~\cite{MisawaPRL,GeorgesHund,Haule}
and superconductivity~\cite{MisawaNcom}.

{In the alkali-doped fullerides $A_{3}$C$_{60}$ (A=K, Rb, Cs), 
low-energy physics is also described by degenerate orbitals --- 
each alkaline atom donates one electron to the three-fold 
degenerate lowest unoccupied molecular orbital 
with the $t_{1u}$ symmetry at the fullerene atom. 
Thus the three $t_{1u}$ orbitals become half filled on average. 

The unique feature of the solid $A_{3}$C$_{60}$ in comparison 
to the above cuprates and the iron-based superconductors is a small 
band width ($\sim 0.7$ eV) ascribed to small overlap of molecular orbitals 
each at the neighboring fullerene atoms, while the 
largest Jahn-Teller $H_g$ phonon frequency coupled to the $t_{1u}$ orbitals is 
comparable and as large as 0.2 eV. It makes the role of phonons substantially large.  
The discovered superconductivity with the highest $T_{\rm c} \sim 40$K is believed 
to have the isotropic $s$-wave symmetry supporting 
the crucial role of phonons~\cite{GunnarssonRMP,GaninNMat,TakabayashiScience,Ganin2010Nature,Zadik2015Science,Mitrano_Nature2016}.

{The superconductivity is, however, found next to the Mott insulator,} 
suggesting the role of electron correlation as well.  
In addition, the superconductivity is found when the three orbitals are half-filled, 
in sharp contrast to the {cuprates} and iron-based superconductors, 
where typically the carrier doping into half filled 
bands is required for the superconductivity.

{It was proposed that the relatively strong Jahn-Teller electron-phonon 
coupling in the alkali-doped fullerides leads to an effectively 
negative (inverted) Hund's rule coupling (IHRC)~\cite{CaponeScience,Capone_RMP2009}. 
By analyzing a degenerate three-orbital model
with the IHRC with the
dynamical mean-field theory (DMFT),
Capone {\it et al.} showed that 
the IHRC 
actually induces the isotropic $s$-wave 
high-$T_{c}$ superconductivity~\cite{CaponeScience}.
Recently, 
quantitative evaluation of the 
interaction parameters including 
the IHRC
from {\it ab initio} calculations has been done~\cite{NomuraScience}.
By solving the obtained {\it ab initio} Hamiltonian 
with the extended DMFT (E-DMFT),
calculated critical temperatures were shown to be 
consistent with the experimental results.}

{In general, superconductors with {$T_{c}$ relatively high} 
in the ratio to the {energy scale of} electron band width 
is found {exclusively} in strongly correlated systems and their small coherence lengths require 
serious account of spatial correlation effects, while DMFT does not consider them. 
Furthermore, severe competitions with magnetic and charge orders 
known in the cuprates and the iron-based superconductors 
urge us to seriously examine the spatial correlation and fluctuation effects.}

{In this Letter, 
we employ the many-variable 
variational Monte Carlo (mVMC) method~\cite{TaharaVMC,misawaHubbard,mVMC}
to 
take account of both spatial and quantum fluctuations.
We first analyze the two-dimensionally projected 
{\it ab initio} Hamiltonian
for the face-centered-cubic 
alkali-doped fullerides (fcc-Cs$_{3}$C$_{60}$).
Then, to capture the essence of the superconductivity, 
we also analyze
simplified three- and two-orbital models {on triangular and square lattices, respectively,}
in which each orbital has transfer
in only one direction and forms one-dimensional chains in each triangular bond direction
and they interact with {each other} only at the same site.}

We reveal the conditions for enhancing superconductivity: 
First a realistic unique three-orbital structure, where the electronic transfer 
at each nearest-neighbor atomic bond is governed by only 
one orbital-diagonal transfer, depending on the bond direction, 
is 
important to reproduce the $s$-wave superconductivity in the realistic parameter region.
The superconductivity is 
replaced by the excitonic Mott insulator with antiferromagnetic order in the strong 
coupling region near the {\it ab initio} parameters and 
replaced by excitonic insulator {without antiferromagnetism} in the two-orbital model. 
The 
superconductivity is universally stabilized 
{through the melting of the excitonic insulators realized by the carrier doping}.
The excitonic insulator is the mother state of the superconductivity 
in the multi-orbital models with the IHRC
and offers the clue for obtaining higher-$T_{c}$ superconductivity
in fullerides and/or other multi-orbital systems with the IHRC.

\begin{figure}[t!]
  \begin{center}
    \includegraphics[width=8.5cm,clip]{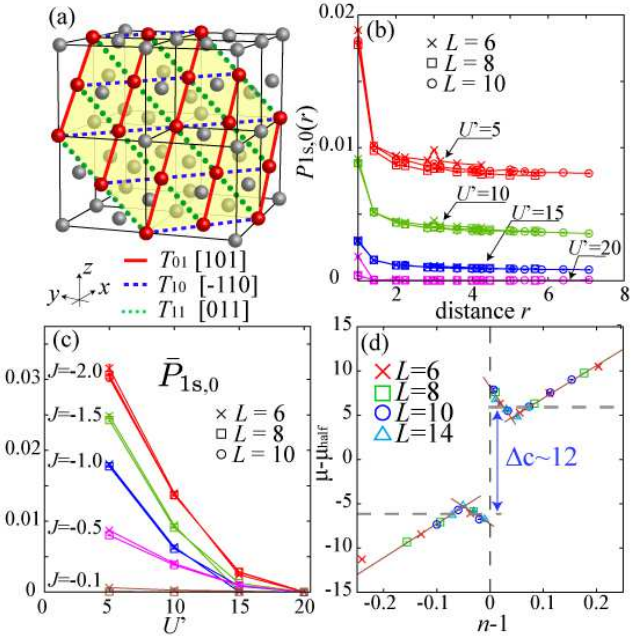}
  \end{center}
\caption{(color online).
(a)~{Two-dimensional projection of the fcc lattice.} 
(b)~Distance dependence of the superconducting correlation $P_{\rm 1s,\nu}(r)$
for several different inter-orbital Coulomb interactions ($U^{\prime}=5,10,15,20$) and
$J=-0.5$. Superconducting correlations do not depend on orbitals.
(c)~Inter-orbital Coulomb interaction $U^{\prime}$ dependence of the 
averaged superconducting correlations $\bar{P}_{\rm 1s,0}$
for several $J$.
(d)~Doping dependence of the chemical potential $\mu$
for $U'=20$ and $J=-0.5$. 
Here, $\mu_{\rm half}$ is the chemical potential at half filling .
The amplitude of the charge gap $\Delta_{c}$ is 
denoted by arrow.
{Because the first-order phase transitions between orbital-polarized 
phase and the superconducting  phase occurs 
at finite doping region~(see \cite{SM}, S.2), 
$\mu$ has kinks around the first-order phase transition~\cite{Ohgoe_2017}.
To perform the Maxwell's construction, 
we fit $\mu$ by linear functions shown by the lines.
Then, by using the fitted linear functions, 
we perform the Maxwell's construction.
Broken lines shows the results of the Maxwell's construction.}
}
\label{fig:phase}
\end{figure}

{\it Model and Method.}--We first study 
three-orbital low-energy $ab$ $initio$  Hamiltonian for
{triangular-lattice cross section of } fcc-fullerides, which is defined by 
\begin{eqnarray}
&H&=\sum_{i,j,\nu,\sigma}t_{ij}(c^{\dagger}_{i\nu\sigma}c_{j\nu\sigma}+{\rm h.c.}) 
+U\sum_{i,\nu}n_{i\nu\uparrow}n_{i\nu\downarrow} \notag \\
&+&U^{\prime}\sum_{i,\nu>\mu}n_{i\nu}n_{i\mu}  
+J_{\rm ex}\sum_{i,\nu<\mu,\sigma,\tau}
c_{i\nu\sigma}^{\dagger}c_{i\mu\tau}^{\dagger}c_{i\nu\tau}c_{i\mu\sigma}
\notag \\
&+&
J_{\rm pair}\sum_{i,\nu<\mu,\sigma,\tau}
c_{i\nu\sigma}^{\dagger}c_{i\nu\tau}^{\dagger}c_{i\mu\tau}c_{i\mu\sigma},
\label{Hamiltonian}
\end{eqnarray}
where $c_{i\nu\sigma}^{\dagger}$ ($c_{i\nu\sigma}$) is the creation (annihilation)
operator on the $\nu$th orbital with spin $\sigma$ at $i$th site and
$n_{i\nu\sigma}=c_{i\nu\sigma}^{\dagger}c_{i\nu\sigma}$ and
$n_{i\nu}=n_{i\nu\uparrow}+n_{i\nu\downarrow}$ are the number operator.

The transfer integrals $t_{ij}$ in the fullerides are given 
by 3$\times$3 matrices~\cite{NomuraPRB}. 
Although the $ab$ $initio$ Hamiltonian is defined on the 
fcc lattice, to reduce the numerical cost,
we project fcc lattice into the triangular lattice {[see Fig.~1(a)]}, i.e.,
we only consider [1,0,1], [-1,1,0], [0,1,1] vectors in the fcc lattice 
and regard them as [1,0], [0,1], [1,1] vectors in the triangular lattice.
(The triangular lattice is represented 
by mapping to the square lattice with diagonal bonds in one direction.) 
In this study, we take only nearest-neighbor transfer 
integrals (in the original triangular lattice representation) because the amplitudes of 
the next-nearest-neighbor transfer integrals are small.
The matrices of the hopping integrals are given as
\begin{align}
T_{01}=
\begin{pmatrix}
F_{3} & 0     & F_{2} \\ 
0     & F_{4} & 0 \\ 
F_{2} & 0     & F_{1}  \\
\end{pmatrix},
T_{10}=
\begin{pmatrix}
F_{1}  & -F_{2} & 0 \\ 
-F_{2} & F_{3}  & 0 \\ 
0      & 0      & F_{4}  \\
\end{pmatrix},
T_{11}=
\begin{pmatrix}
F_{4} & 0     & 0 \\ 
0     & F_{1} & F_{2} \\ 
0     & F_{2} & F_{3}  \\
\end{pmatrix}.
\end{align}
{In this study, we use the {\it ab initio} parameters for
fcc-Cs$_3$C$_{60}$, {which shows} maximum $T_{c}$, denoted 
by fcc-Cs($V_{\rm SC}^{\rm opt-S}$) in the literature~\cite{NomuraPRB}.
In fcc-Cs($V_{\rm SC}^{\rm opt-S}$), $F_{3}$
is 
0.0372 eV and we take this value as energy unit.
Other hopping parameters are given as
$F_{1}/|F_{3}|\sim 0.07$,
$F_{2}/|F_{3}|\sim-0.80$,
$F_{4}/|F_{3}|\sim-0.32$.}

{According to the $ab$ $initio$ calculations~\cite{NomuraScience},
interaction parameters that is renormalized by the
electron phonon interactions
are estimated as
$U^{\prime}/|F_{3}|\sim22.20$,
$U/|F_{3}|\sim 23.04$, and
$J=J_{\rm ex}=J_{\rm pair}$ with $J/|F_{3}|\sim -0.43$.
They nearly satisfies the rotational symmetry, i.e.,
$U=U^{\prime}+2J$.
In our calculations, we systematically monitor
the interaction parameters with the 
constraint  $U=U^{\prime}+2J$ 
around the above realistic parameter region.
{We take $N_{\rm all}=N_{\rm s} \times N_{\rm orb}$ sites with 
periodic-periodic boundary conditions, 
where $N_{\rm s}=L\times L $ ($N_{\rm orb}$) denotes number of sites (the number of orbitals).
We define the electron density as 
$n=\sum_{i,\nu,\sigma}n_{i\nu\sigma}/N_{\rm all}$.}
}

In the mVMC method~\cite{TaharaVMC,misawaHubbard,mVMC},
the variational wave function is defined as
$|\psi\ra =\sP_{\rm G}\sP_{\rm J}|\phi_{\rm pair}\ra$,
where $\sP_{\rm G}$ and $\sP_{\rm J}$
are the
{Gutzwiller factors~\cite{Gutzwiller} defined as
  $\sP_{\text{G}} = \exp(-\sum_{i,\nu}g_{\nu}n_{i\nu\uparrow}n_{i\nu\downarrow})$ and
the Jastrow factors~\cite{Jastrow,CapelloJastrow},
defined as 
$\sP_{\text{J}} = \exp(-\frac{1}{2} \sum_{i,j,\nu,\mu} v_{ij\nu\mu} n_{i\nu} n_{j\mu})$,
respectively.}
{The 
{pair-product} part $|\phi_{\rm pair}\ra$ is
the generalized pairing wave function defined as
$|\phi_{\rm pair}\ra= \Big[\sum_{i,j=1,\nu,\mu}^{\Ns}f_{ij\nu\mu}c_{i\nu\uparrow}^{\dag}c_{j\mu\downarrow}^{\dag}\Big]^{N_{e}/2} |0 \ra$,
where $f_{ij}$ denotes the variational parameters.
For details of {wave functions}, 
see Refs.~\onlinecite{gros1989physics,BajdichPRB,TaharaVMC,mVMC}.
In this Letter, 
we have $2\times2\times N_{\rm orb} \times N_{\rm s}$
independent variational parameters for {pair-product} part.
All the variational parameters are simultaneously
optimized by using the stochastic
reconfiguration method~\cite{Sorella_PRB2001,TaharaVMC}.
The variational function $|\psi\ra$
can flexibly describe several phases such as 
correlated paramagnetic metals, 
antiferromagnetic/charge-ordered phases
and superconducting phases
as well as their coexistence.}

{{\it Results.}--} {First, we examine the stability of the superconducting phase
at half filling
in the two-dimensionally projected 
$ab$ $initio$ Hamiltonian (we call this model full model).
To detect the superconductivity,
we calculate the isotropic $s$-wave superconducting correlations, 
which is defined as 
$P_{\rm 1s,\nu}(\vec{r})={1}/{2N_{\rm s}}
\sum_{i=1}^{N_{\rm s}}\langle 
\Delta^{\dagger}_{\rm 1s,\nu}(\vec{r_{i}})\Delta_{\rm 1s,\nu}(\vec{r_{i}}+\vec{r}) 
+\Delta_{\rm 1s,\nu}(\vec{r_{i}})\Delta^{\dagger}_{\rm 1s,\nu}(\vec{r_{i}}+\vec{r}) 
\rangle$, where $\Delta_{\rm 1s,\nu}(\vec{r}_{i})={1}/{\sqrt{2}}(c_{i\nu\uparrow}c_{i\nu\downarrow}-c_{i\nu\downarrow}c_{i\nu\uparrow})$.
We also calculate the average value of $P_{\rm 1s,\nu}(\vec{r})$
at long distance ($|\vec{r}|\geq4$), which is denoted 
by $\bar{P}_{\rm 1s,\nu}$.
As shown in Fig.~1(b),
we find that the superconducting correlations
{approach a nonzero} constant value at long distance 
for $J=-0.5, U^{\prime}\leq 15$. 
{The size dependence of the saturated value is small and
this indicates that the 
superconducting phase {is} stable 
for $U^{\prime}\leq {13}$.
As we show later, insulating phase competes with superconducting phase and
superconductor-insulator transition occurs around $U^{\prime}\sim13$.}

Next, we systematically 
study the interaction dependence. As shown in Fig.~1(c),
the superconducting correlations
are largely enhanced by increasing $J$ in the moderately strong coupling
region ($U^{\prime}\leq 10$). 
{This result indicates that the superconductivity is induced by the 
Suhl-Kondo mechanism~\cite{Suhl_PRL1959,Kondo_PTP1963}.
The 
enhancement is mainly induced by $J_{\rm pair}$~(see \cite{SM}, S.1).}

{To analyze the nature in the strong coupling region,
the doping dependence of the 
chemical potential defined by
$\mu(\bar{N})=\{E(N_{1})-E(N_{2})\}/\{N_{1}-N_{2}\}$, {which} {is shown in Fig.~1(d). Here,} 
$E(N_{1})$ is the total energy at filling $N_{1}$ and
$\bar{N}=(N_{1}+N_{2})/2$.
The insulating phase {indicated by the gap $\Delta_c$ in $\mu$} appears in the strong coupling region.
}

\begin{figure}[t!]
  \begin{center}
    \includegraphics[width=8.5cm,clip]{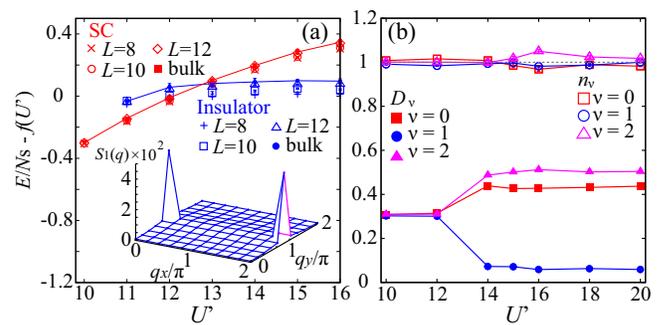}
  \end{center}
\caption{(color online).
{(a)~Inter-orbital Coulomb interaction 
$U^{\prime}$ dependence of the energy for {the 
superconducting and insulating phases.}
We subtract the linear function $f(U^{\prime})$ from the 
energies for better clarity.
We estimate the thermodynamic limit of the energy
by assuming the relation $E(L)/L^2=E(L=\infty)+a_{1}/L^3$,
where $E(L)$ denotes the energy for $L\times L$ system size.
The first-order phase transition occurs at $U\sim12.8$.
In the inset, the spin structure factor 
in the insulating phase ($L=10,U^{\prime}=20,J=-0,5$) is shown.
The spin structure factor is defined as
$S_{\nu}(\bm{q})={1}/{3N_{s}^2}\sum_{i,j}
\langle \boldsymbol{S}_{i\nu}\cdot\boldsymbol{S}_{j\nu}\rangle e^{i\bm{q}\cdot(\bm{r}_{i}-\bm{r}_{j})}$,
where 
$\bm{S}_{i}=1/2\sum_{\sigma,\sigma^{\prime}}c^{\dagger}_{i,\sigma}\boldsymbol{\sigma}_{\sigma,\sigma^{\prime}}c_{i,\sigma^{\prime}}$ and 
$\boldsymbol{\sigma}$ denotes the Pauli matrices. 
}
(b)~Inter-orbital Coulomb interaction 
$U^{\prime}$ dependence of the orbital occupation $n_{\nu}$
and orbital-dependent doublon density $D_{\nu}$ ($J=-0.5$ and $L=10$).
In the superconducting phase, the doublon density does not depend on orbitals while 
this symmetry is broken in the insulating phase.
}
\label{fig:Ins}
\end{figure}

{
In Fig.~2(a), we show $U^{\prime}$ dependence of energies
for {the superconducting and insulating phases.
The energy crossing indicates
the first-order phase transition} at $U^{\prime}\sim12.8$.
Because the {\it ab initio} interaction parameter
is estimated as $U^{\prime}\sim 22$~\cite{NomuraScience},
{the insulating phase looks overestimated on the triangular lattice. However, this is 
naturally understood from the bandwidth ($W$) of the triangular lattice, 
which is a half of the fcc lattice. 
Since the ratio $U/W$ primarily determines 
the metal (superconductor)-insulator transition,
the transition found at $U^{\prime}=12.8$ here corresponds
to $U^{\prime}=25.6$ for the fcc lattice. 
Therefore the {\it ab initio} value $U^{\prime}\sim22$ is close to
the transition but in the superconducting 
phase in agreement with the experimental result.
On the other hand, we note that
IHRC is required to keep a value 
comparable to the {\it ab initio} value $\sim0.5F_{3}$ to
stabilize the superconductivity even for 
the present model on  the two-dimensional
triangular lattice.
}

{To further examine the nature of the insulating phase,
we show the orbital occupancies and orbital-dependent
doublon densities as a function of $U^{\prime}$ in Fig.~2(b),
which are defined as 
$n_{\nu}=1/N_{\rm s}\sum_{i,\sigma}n_{i\nu\sigma}$ and
$D_{\nu}=1/N_{\rm s}\sum_{i,\sigma}n_{i\nu\uparrow}n_{i\nu\downarrow}$.
$D_{\nu}$ 
{shows orbital differentiation and thus the orbital symmetry breaking} in the insulating phase.}

In this orbital-differentiated Mott insulator, two orbitals are disordered and the antiferromagnetic order exists only in one orbital
with the small doublon density [see the inset in Fig.~2(a) and S.~3 in \cite{SM}].
{Because $\nu=0$ orbital has
larger transfer in $[01]$ direction ($F_{4}$) compared with
$\nu=2$ orbital, short range antiferromagnetic 
correlation is induced by 
the proximity effect of the antiferromagnetic
order in $\nu=1$ orbital [see S.~4 in \cite{SM}]. 
This short-range antiferromagnetic correlation induces the
small but finite differences 
in $D_{\nu}$ between the disordered orbitals ($\nu=0$ and $\nu=2$).}

We further note that a genuine 
orbital-differentiated Mott insulator without any {spin} symmetry breaking {exists}
as a low-energy excited state and it's  
{energy relative to the antiferromagnetic ground state} is very small ($\sim 0.04 F_3$) (see \cite{SM}, S.5).
This near degeneracy of the magnetically ordered and disordered states
is consistent with the experimental results where
the antiferromagnetic transition occurs at a
very low temperature for fcc-Cs$_3$C$_{60}$ ($T_{\rm N}\sim 2$K)~\cite{Ganin2010Nature};
its ordered moment is considerably small and 
it was suggested that antiferromagnetic phase coexists with the
magnetically disordered phase~\cite{KasaharaPRB2014}.
{We will discuss the nature of the 
orbital-differentiated Mott insulator later.}

Next, to capture the essence of the superconductivity {and the insulating phase} found in the
full model, we {study a simplified model.} 
One characteristic feature of the transfer matrices in the full model is
that the directions of the largest hopping $F_{3}$ depend on the orbitals. 
If we consider only the largest hopping $F_{3}$,
the three orbitals form three chains as shown in Fig.~1(a).
In this model, {if one maps the triangular lattice to the square lattice with the diagonal hopping in one direction [11],} each orbital has hopping
only in {either} [10], [01], or [11] directions, depending on the orbitals.
{Electrons on different chains} interact only at the same site through $U$, $U^{\prime}$, and $J$ terms.}
We call this simplified model the $F_{3}$ model.

\begin{figure}[t!]
  \begin{center}
    \includegraphics[clip,width=8.5cm]{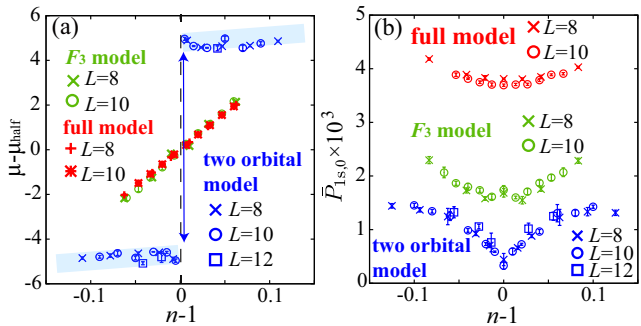}
  \end{center}
\caption{(color online).
(a)~Doping dependence of 
the chemical potential in {the three models at $J=-0.5$ and $U^{\prime}=10$}.
In the two-orbital model, the charge gap {opens}.
Smooth doping dependences of  $\mu$ in the $F_{3}$ model 
and the full model are evidence of {the absence of the insulator  
at half filling, which turns out to be the superconducting state in (b).}
(b)~Doping dependence of the 
averaged superconducting correlations
in {the three models at$J=-0.5$ and $U^{\prime}=10$}. 
}
\label{fig:SC}
\end{figure}

{
To reveal the multi-orbital effects on the superconductivity,
we also consider a two-orbital model on the square lattice
analogously to the  
$F_{3}$ model. 
In the two-orbital model,
the orbital {0 (1)} has hopping 
only in the [10] ([01]) direction.}
{The $F_{3}$ and two orbital} 
models {have some resemblance to} the Kitaev model~\cite{Kitaev} in the sense that
they have orbital-dependent spatial anisotropy. 

{In the case of the two-orbital model, insulating charge gap opens for any positive $U^{\prime}$ as shown in an example in Fig.~3(a).
The nature of the insulator is understood as
a doublon and a holon locally 
bound and resonating in the form of Frenkel excitons at each site with the degenerate two orbitals, 
while they do not break any translational symmetry, }  
{where charge/spin structure factors
do not have appreciable peaks (see \cite{SM}, S.6)}.
This local configuration indeed does not have
energy loss of the interorbital Coulomb interaction $U^{\prime}$.
Therefore, the ground state of the two-orbital model is 
interpreted as an excitonic insulator phase without any spatial symmetry breaking and
stabilized by electron (originally including electron-phonon) 
correlations where the orbitals of doubly occupied (doublon) and empty (holon) orbitals
do not spatially order, but is represented by the 
linear combination $|d_1h_2\rangle \pm |d_2h_1\rangle$ at each site, where for 
instance $d_1(h_2)$ expresses the doublon (holon) at the 
orbital 1 (2). 

{By doping carriers into the {excitonic} {(bosonic)} Mott insulator,
the superconducting phase 
immediately appears as shown in Fig.~3(b), 
where doublons (local Cooper pair) condense. 
Although the local Cooper pair is already formed at half filling, it is frozen.}

{{\it Discussion and Summary.}--}
{In the three-orbital models {(full and $F_{3}$ models)},
because {the three electrons occupy a site at half filling at large $U$ and $U'$, 
an exciton (one doublon and one holon) formation leaves one electron unpaired at the third orbital.
This introduces much larger charge fluctuations than the two-orbital model. }
We find that
this {charge fluctuation easily causes charge melting of the excitonic Mott insulator phase found} 
in the two-orbital model and
the superconductivity appears even at half filling.

{If the local repulsions $U$ and $U'$ become sufficiently strong,} 
{in the full model,
one unpaired orbital finally becomes the antiferromagnetic Mott insulator
and other orbitals with Frenkel exciton forms 
an excitonic insulator {similarly to} the two orbital model. 
This phase may be called an orbital-differentiated Mott insulator. 
{A similar phase in a three-orbital model was discussed before}~\cite{HoshinoPRL}.
}

{In the three-orbital 
model, a disordered state exists as a low-energy excited state in the strong coupling region, where
spin order is absent.
Since the averaged electron filling per unit cell is odd (three),
this spin-disordered state is a genuine Mott insulator, 
which is not adiabatically connected to the band insulator.
In other words, this phase is a promising candidate of 
quantum spin liquid if its energy can be lowered below the antiferromagnetic state and
further detailed study of tuning is an intriguing issue
but left for future studies.
}

{A$_3$C$_{60}$ forms a fcc or A15 structure 
instead of the 2D lattice considered here. We expect that the essential physics is 
the same because each bond has basically only one dominant 
diagonal hopping $F_{3}$ and the same excitonic Mott insulator is likely located nearby. 
Detailed study on the real lattice structure is an important future issue. }

{In summary, {
for the two-dimensional version of the $ab$ $initio$ Hamiltonian
(full model) for the alkaline doped fullerene, we have found that the isotropic $s$-wave superconducting phase becomes the ground state in the realistic parameter region.
In stronger coupling region, 
a superconductor-insulator phase 
transition occurs, where the insulating phase is
interpreted as {an orbital differentiated Mott insulator consisting of a two-orbital excitonic Mott insulator and a one-orbital antiferromagnetic insulator.
The antiferromagnetic insulator in the third orbital is nearly degenerate with the spin-liquid phase.}
To extract the essence, we have also analyzed simple 
three- and two-orbital models,
consisting of network of chains with orbital-dependent anisotropic transfer.
The comparison of the two- and
three-orbital models suggests that 
the melting of the excitonic insulator by the carrier doping 
or by introducing the third frustrating orbital
stabilize isotropic $s$-wave superconductivity.
Although the origin of the superconductivity is ascribed to
the strong attractive force arising from the IHRC,
the attractive force generally induces the insulating phase at half filling.
To melt the insulator at half filling, 
the odd number of orbitals is helpful.
Our study on the multi-orbital systems with
the IHRC shows a new route to obtain stable superconductors
as well as a quantum spin liquid in the systems with the IHRCs.}

This work was financially supported by a Grant-in-Aid for 
Scientific Research (No.~22104010, No.~16H06345 and No.~16K17746) 
from Ministry of Education, Culture, Sports, Science and Technology, Japan. 
This work was also supported in part by MEXT as a social and scientific priority 
issue (Creation of new functional devices and 
high-performance materials to support next-generation industries GCDMSI) 
to be tackled by using post-K computer. The authors thank the 
Supercomputer Center, the Institute for Solid State Physics, 
the University of Tokyo for the facilities. We thank the computational 
resources of the K computer provided by the RIKEN 
Advanced Institute for Computational Science through the 
HPCI System Research project (hp150173, hp150211, hp160201,hp170263) 
supported by Ministry of Education, Culture, Sports, Science, and Technology, Japan.
TM is supported by Building of Consortia for the Development 
of Human Resources in Science and Technology from the MEXT of Japan.

\clearpage
\noindent
{\Large
Supplemental  Materials  
}



\renewcommand{\figurename}{Fig. S}
\renewcommand{\theequation}{S.\arabic{equation}}
\setcounter{equation}{0}
\setcounter{figure}{0}
\renewcommand{\tablename}{Table S}

\section{S.1~ Pair hopping $J_{\rm pair}$ and 
exchange interaction $J_{\rm ex}$ dependence of the superconducting correlations}
In the main text, we show that the superconducting correlations
are mainly governed by the IHRC.
{The IHRC consists of
two parts, i.e., the exchange Hund's rule coupling $J_{\rm ex}$,
and the pair hopping term $J_{\rm pair}$ in Eq. (1).
As shown in Fig.S~\ref{fig:depJ}, we show that  
superconducting correlations are largely suppressed by decreasing
the amplitude of the pair hopping terms 
and it becomes nearly zero for $J_{\rm pair}=0$.
In contrast to that,
$J_{\rm ex}$ does not alter the superconducting 
correlations.
This result clearly shows that the pair 
hopping is the crucial interaction for the superconductivity.
We note that 
the previous studies with DMFT for
similar multi-orbital systems claims
that superconductivity appears 
without pair hoppings~\cite{KogaPRB,HoshinoPRL}.
In the light of our result, these results 
should be reexamined whether they are artifacts of the DMFT 
because the DMFT generally overestimate the superconductivity
due to its mean-field nature.}

\begin{figure}[hb!]
  \begin{center}
    \includegraphics[width=7cm,clip]{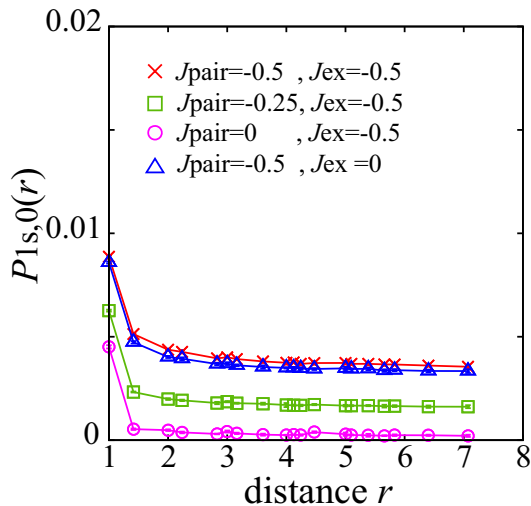}
  \end{center}
\caption{(color online)~Pair hopping $J_{\rm pair}$ and exchange interaction $J_{\rm ex}$
dependences of the superconducting correlations for $U^{\prime}=10$, $U=9$, and $L=10$.
We note that the relation $U=U^{\prime}+2J$ is not satisfied in this calculation.
}
\label{fig:depJ}
\end{figure}

\newpage
\section{S.2~Doping dependence of the orbital-dependent
doublon densities $D_{\nu}$}
We show the doping dependence of the 
orbital-dependent doublon densities $D_{\nu}$ in Fig.S~\ref{fig:DopeD}.
Around half filling, the spontaneous symmetry breaking  of orbital still
exists and it {vanishes at large doping region,} where 
the superconducting phase becomes stable.
A first-order phase transition between the excitonic phase with orbital-differentiated doublons and the superconducting phase occurs at finite doping.
The doping dependence of the 
chemical potential has kink structure as 
we mentioned in the main text.

\begin{figure}[hb!]
  \begin{center}
    \includegraphics[width=7cm,clip]{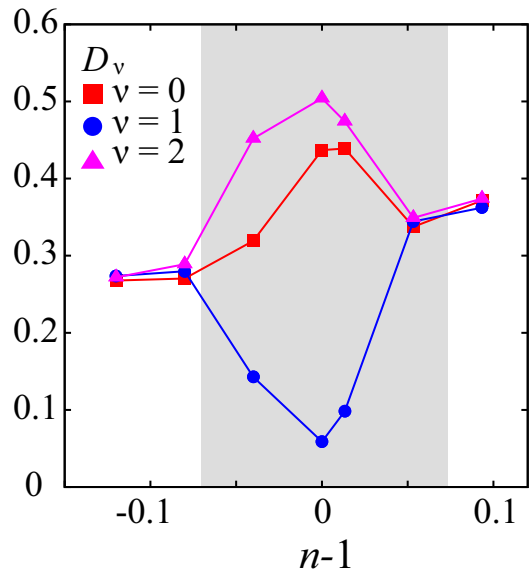}
  \end{center}
\caption{(color online)~
Doping dependence of $D_{\nu}$ for $U'=20$, $J=-0.5$, and $L=10$.
Shaded region shows the phase separation determined from the
Maxwell's construction [see main text and caption in Fig.S~1(d)].
}
\label{fig:DopeD}
\end{figure}

\newpage
\section{S.3~Size dependence of the spin structure factors in the full model}
In Fig.S~\ref{fig:SqAll}, we show the spin structure factors
in the strong coupling region
for several different systems sizes.
For small system sizes ($L=6,8$), we find that the spin structure factors
have peak at $\vec{q}=(\pi,\pi)$ while they have peak 
at  $\vec{q}=(0,\pi),(\pi,0)$ in large system sizes ($L=10,12$).
Although it is hard to perform calculations for larger system sizes,
it is plausible that stripe magnetic order ($\vec{q}_{\rm peak}=(0,\pi),(\pi,0)$) 
becomes stable in the thermodynamic limit. 

The size extrapolation 
of the spin structure factors is shown in Fig.S~\ref{fig:ext}.
Although the positions of the Bragg peaks depend on
the system sizes, the size dependence of the peak value is smooth.
We estimate the thermodynamic value of the spin structure factors
as $S_{1}(\vec{q}_{\rm peak})=0.041(4)$.

\begin{figure}[hb!]
  \begin{center}
    \includegraphics[width=9cm,clip]{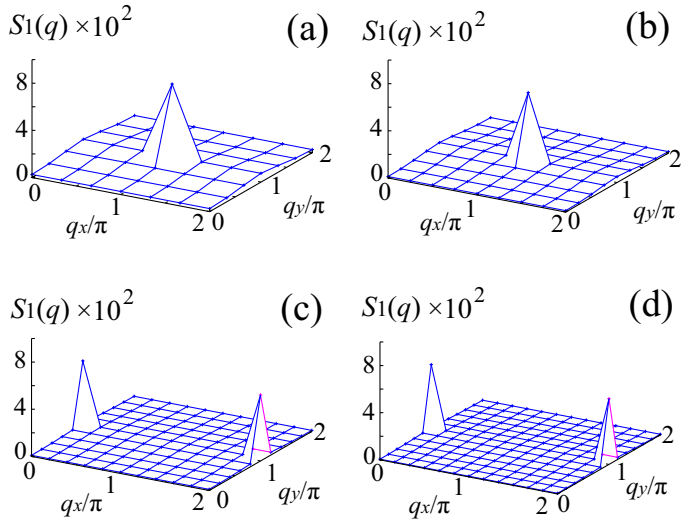}
  \end{center}
\caption{(color online)~
Spin structure factors for several different system sizes
(a)$L=6$, (b)$L=8$, (c)$L=10$, and (d)$L=12$.
We take $U^{\prime}=20$ and $J=-0.5$.
}
\label{fig:SqAll}
\end{figure}

\begin{figure}[ht!]
  \begin{center}
    \includegraphics[width=7cm,clip]{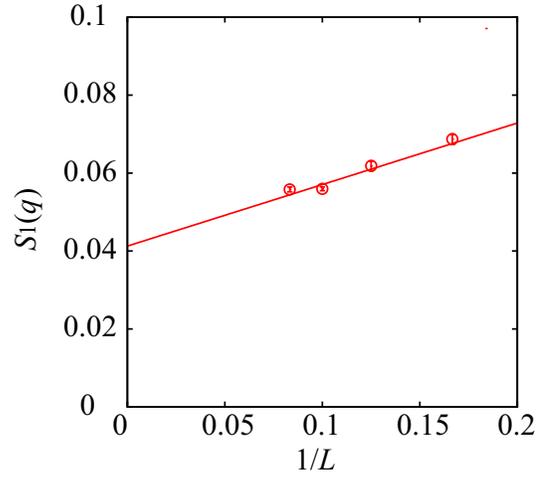}
  \end{center}
\caption{(color online)~
Size extrapolation of the spin structure factor.
}
\label{fig:ext}
\end{figure}

\newpage
\clearpage
\section{S.4~Short-range antiferromagnetic correlations of disorders orbitals in the full model}
{In Fig.S~\ref{fig:short}, we show
 how the antiferromagnetic order in $\nu=1$ orbital
affects the spin structure 
factors on the other disordered orbitals $\nu=0$ and $\nu=2$.
Because $\nu=0$ orbital has relatively larger hopping in $[01]$ direction
compared to $\nu=2$ orbital, 
$\nu=0$ orbital is largely
affected by the antiferromagnetic order
and shows short-range antiferromagnetic order
as shown in Fig.S~\ref{fig:short}(a). }

\begin{figure}[htb!]
  \begin{center}
    \includegraphics[width=7cm,clip]{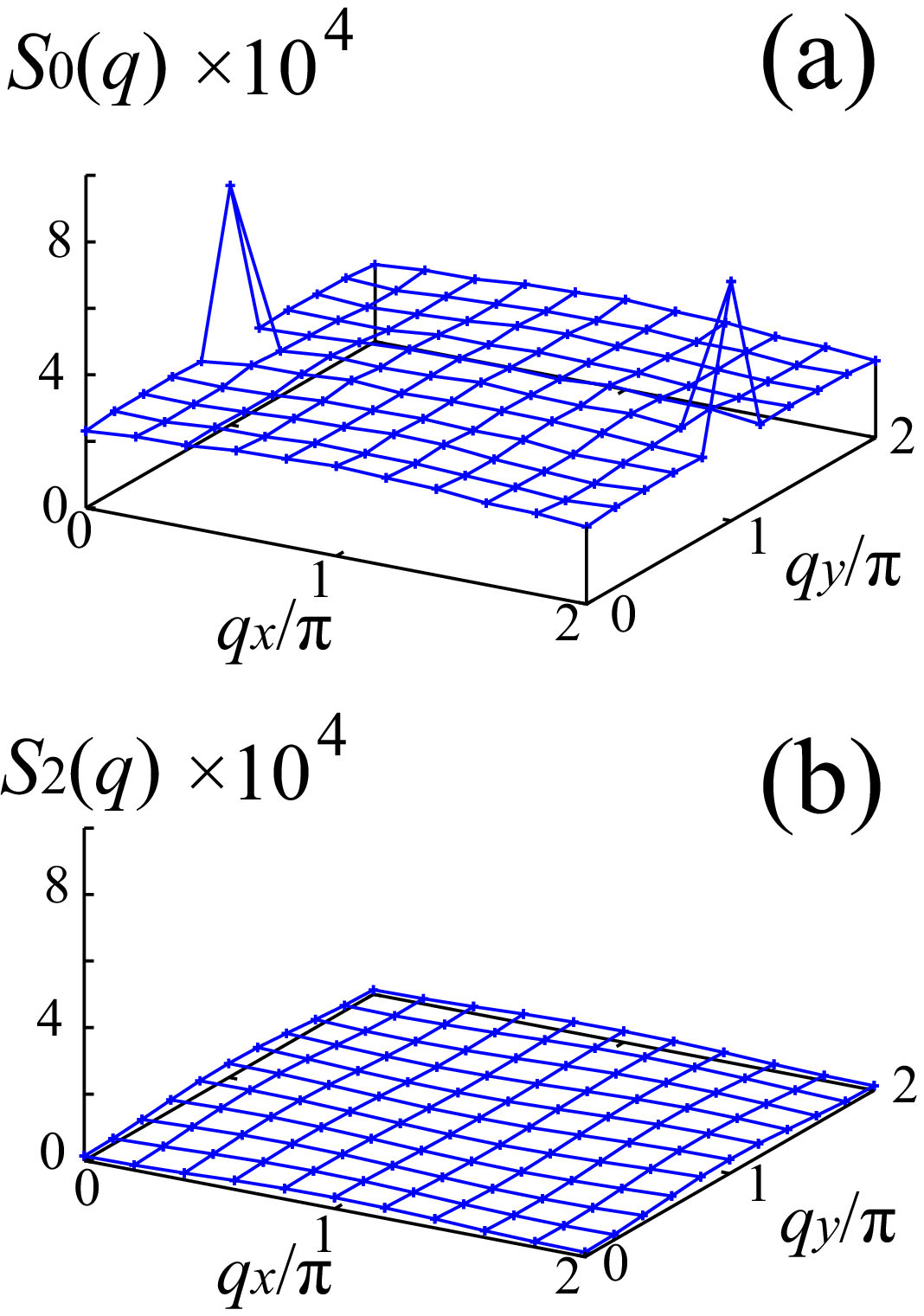}
  \end{center}
\caption{(color online)~
Spin structure factors [(a)$\nu=0$ and (b)$\nu=2$] 
for $U^{\prime}=20$, $J=-0.5$, and $L=10$.
}
\label{fig:short}
\end{figure}

\clearpage
\section{S.5~Low-energy excited state in the full model}
Here, we mention the low-energy excited state in the insulating 
phase of the full model.
For $L=8$, the low-energy excited state without spin/charge 
order. The spin and charge structure factors are shown in Fig.S~\ref{fig:exc}.
The charge structure factors is defined as 
\begin{align}
N_{\nu}({\vec{q}})=\frac{1}{N_{\rm s}^2}\sum_{i,j}n_{i\nu}n_{j\nu}e^{i\vec{q}(\vec{r}_i-\vec{r}_{j})}.
\end{align}
We do not find any significant peaks in the spin and 
charge structure factors.
Its energy {per site} is $E_{\rm exc}=58.645(1)$ (energy unit is $F_{3}\sim 372 {\rm K}$)
while the ground state energy is 
$E_{\rm g}=58.599(2)$.
The energy difference $\Delta E$ is given by 
$\Delta E\sim 0.04 \sim 15 {\rm K}$.
This nearly degenerate state 
may make the Neel temperature low and
induce the peculiar coexistence of the magnetic order and
the disordered state observed in experiment~\cite{KasaharaPRB2014}.

\begin{figure}[htb!]
  \begin{center}
    \includegraphics[width=8cm,clip]{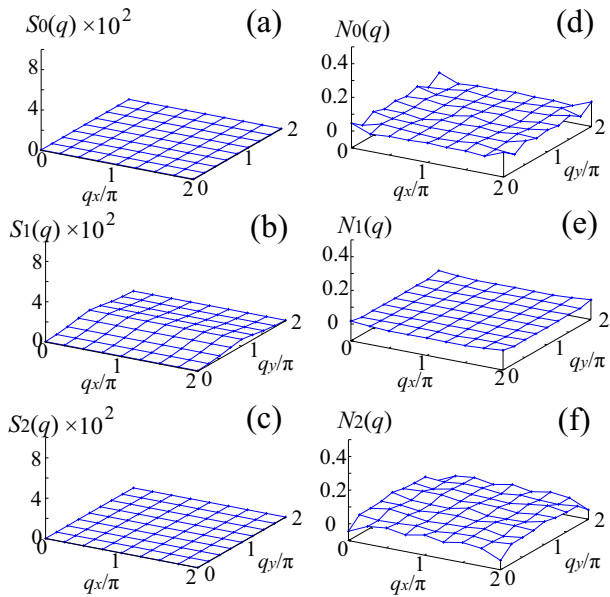}
  \end{center}
\caption{(color online)~
Spin structure factors [(a)$\nu=0$, (b)$\nu=1$, and (c)$\nu=2$] 
and charge structure factors [(d)$\nu=0$, (e)$\nu=1$, and (f)$\nu=2$]  
in the low-energy excited state of the full model for
 $U^{\prime}=20$, $J=-0.5$, and $L=8$.
}
\label{fig:exc}
\end{figure}

\clearpage
\section{S.6~Spin and charge structure factors in the two-orbital model}
The spin and charge structure factors are shown 
in Fig.S~\ref{fig:two}.
We do not find any significant peak in the 
structure factors.
We note that orbital differentiation of 
doublons does not occur in the two-orbital model, i.e., 
$D_{0}=0.487(7)$ and $D_{1}=0.487(4)$ 
for the same parameters in Fig.S~\ref{fig:two}.

\begin{figure}[hb!]
  \begin{center}
    \includegraphics[width=9cm,clip]{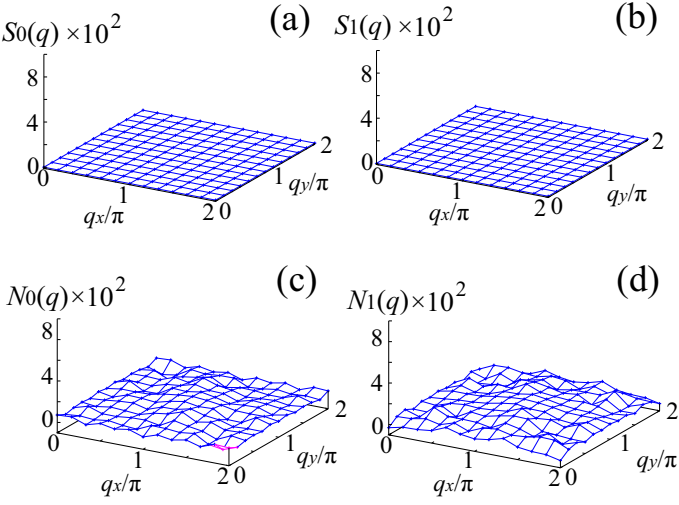}
  \end{center}
\caption{(color online)~
Spin structure factors [(a)$\nu=0$ and (b)$\nu=1$] and
charge structure factors [(c)$\nu=0$ and (d)$\nu=1$] in the two-orbital model for
$U^{\prime}=10$, $J=-0.5$ and $L=12$.
}
\label{fig:two}
\end{figure}

\begin{thebibliography}{35}
\expandafter\ifx\csname natexlab\endcsname\relax\def\natexlab#1{#1}\fi
\expandafter\ifx\csname bibnamefont\endcsname\relax
  \def\bibnamefont#1{#1}\fi
\expandafter\ifx\csname bibfnamefont\endcsname\relax
  \def\bibfnamefont#1{#1}\fi
\expandafter\ifx\csname citenamefont\endcsname\relax
  \def\citenamefont#1{#1}\fi
\expandafter\ifx\csname url\endcsname\relax
  \def\url#1{\texttt{#1}}\fi
\expandafter\ifx\csname urlprefix\endcsname\relax\def\urlprefix{URL }\fi
\providecommand{\bibinfo}[2]{#2}
\providecommand{\eprint}[2][]{\url{#2}}

\bibitem[{\citenamefont{Bednorz and M{\"u}ller}(1986)}]{Bednorz}
\bibinfo{author}{\bibfnamefont{J.~G.} \bibnamefont{Bednorz}} \bibnamefont{and}
  \bibinfo{author}{\bibfnamefont{K.~A.} \bibnamefont{M{\"u}ller}},
  \bibinfo{journal}{Z. Phys.} \textbf{\bibinfo{volume}{64}},
  \bibinfo{pages}{189} (\bibinfo{year}{1986}).

\bibitem[{\citenamefont{Imada et~al.}(1998)\citenamefont{Imada, Fujimori, and
  Tokura}}]{ImadaRMP}
\bibinfo{author}{\bibfnamefont{M.}~\bibnamefont{Imada}},
  \bibinfo{author}{\bibfnamefont{A.}~\bibnamefont{Fujimori}}, \bibnamefont{and}
  \bibinfo{author}{\bibfnamefont{Y.}~\bibnamefont{Tokura}},
  \bibinfo{journal}{Rev. Mod. Phys.} \textbf{\bibinfo{volume}{70}},
  \bibinfo{pages}{1039} (\bibinfo{year}{1998}).

\bibitem[{\citenamefont{Lee et~al.}(2006)\citenamefont{Lee, Nagaosa, and
  Wen}}]{LeeRMP}
\bibinfo{author}{\bibfnamefont{P.~A.} \bibnamefont{Lee}},
  \bibinfo{author}{\bibfnamefont{N.}~\bibnamefont{Nagaosa}}, \bibnamefont{and}
  \bibinfo{author}{\bibfnamefont{X.-G.} \bibnamefont{Wen}},
  \bibinfo{journal}{Rev. Mod. Phys.} \textbf{\bibinfo{volume}{78}},
  \bibinfo{pages}{17} (\bibinfo{year}{2006}).

\bibitem[{\citenamefont{Kamihara et~al.}(2008)\citenamefont{Kamihara, Watanabe,
  Hirano, and Hosono}}]{Kamihara_LaFeAsO}
\bibinfo{author}{\bibfnamefont{Y.}~\bibnamefont{Kamihara}},
  \bibinfo{author}{\bibfnamefont{T.}~\bibnamefont{Watanabe}},
  \bibinfo{author}{\bibfnamefont{M.}~\bibnamefont{Hirano}}, \bibnamefont{and}
  \bibinfo{author}{\bibfnamefont{H.}~\bibnamefont{Hosono}},
  \bibinfo{journal}{J.\ Am.\ Chem.\ Soc.} \textbf{\bibinfo{volume}{130}},
  \bibinfo{pages}{3296} (\bibinfo{year}{2008}).

\bibitem[{\citenamefont{Misawa et~al.}(2012)\citenamefont{Misawa, Nakamura, and
  Imada}}]{MisawaPRL}
\bibinfo{author}{\bibfnamefont{T.}~\bibnamefont{Misawa}},
  \bibinfo{author}{\bibfnamefont{K.}~\bibnamefont{Nakamura}}, \bibnamefont{and}
  \bibinfo{author}{\bibfnamefont{M.}~\bibnamefont{Imada}},
  \bibinfo{journal}{Phys. Rev. Lett.} \textbf{\bibinfo{volume}{108}},
  \bibinfo{pages}{177007} (\bibinfo{year}{2012}).

\bibitem[{\citenamefont{Georges et~al.}(2013)\citenamefont{Georges, Medici, and
  Mravlje}}]{GeorgesHund}
\bibinfo{author}{\bibfnamefont{A.}~\bibnamefont{Georges}},
  \bibinfo{author}{\bibfnamefont{L.~d.} \bibnamefont{Medici}},
  \bibnamefont{and} \bibinfo{author}{\bibfnamefont{J.}~\bibnamefont{Mravlje}},
  \bibinfo{journal}{Annu. Rev. Condens. Matter Phys.}
  \textbf{\bibinfo{volume}{4}}, \bibinfo{pages}{137} (\bibinfo{year}{2013}).

\bibitem[{\citenamefont{Haule et~al.}(2008)\citenamefont{Haule, Shim, and
  Kotliar}}]{Haule}
\bibinfo{author}{\bibfnamefont{K.}~\bibnamefont{Haule}},
  \bibinfo{author}{\bibfnamefont{J.~H.} \bibnamefont{Shim}}, \bibnamefont{and}
  \bibinfo{author}{\bibfnamefont{G.}~\bibnamefont{Kotliar}},
  \bibinfo{journal}{Phys. Rev. Lett.} \textbf{\bibinfo{volume}{100}},
  \bibinfo{pages}{226402} (\bibinfo{year}{2008}).

\bibitem[{\citenamefont{Misawa and Imada}(2014{\natexlab{a}})}]{MisawaNcom}
\bibinfo{author}{\bibfnamefont{T.}~\bibnamefont{Misawa}} \bibnamefont{and}
  \bibinfo{author}{\bibfnamefont{M.}~\bibnamefont{Imada}},
  \bibinfo{journal}{Nat. Commun.} \textbf{\bibinfo{volume}{5}},
  \bibinfo{pages}{5738} (\bibinfo{year}{2014}{\natexlab{a}}).

\bibitem[{\citenamefont{Gunnarsson}(1997)}]{GunnarssonRMP}
\bibinfo{author}{\bibfnamefont{O.}~\bibnamefont{Gunnarsson}},
  \bibinfo{journal}{Rev. Mod. Phys.} \textbf{\bibinfo{volume}{69}},
  \bibinfo{pages}{575} (\bibinfo{year}{1997}).

\bibitem[{\citenamefont{Ganin et~al.}(2008)\citenamefont{Ganin, Takabayashi,
  Khimyak, Margadonna, Tamai, Rosseinsky, and Prassides}}]{GaninNMat}
\bibinfo{author}{\bibfnamefont{A.~Y.} \bibnamefont{Ganin}},
  \bibinfo{author}{\bibfnamefont{Y.}~\bibnamefont{Takabayashi}},
  \bibinfo{author}{\bibfnamefont{Y.~Z.} \bibnamefont{Khimyak}},
  \bibinfo{author}{\bibfnamefont{S.}~\bibnamefont{Margadonna}},
  \bibinfo{author}{\bibfnamefont{A.}~\bibnamefont{Tamai}},
  \bibinfo{author}{\bibfnamefont{M.~J.} \bibnamefont{Rosseinsky}},
  \bibnamefont{and}
  \bibinfo{author}{\bibfnamefont{K.}~\bibnamefont{Prassides}},
  \bibinfo{journal}{Nat. Mat.} \textbf{\bibinfo{volume}{7}},
  \bibinfo{pages}{367} (\bibinfo{year}{2008}).

\bibitem[{\citenamefont{Takabayashi et~al.}(2009)\citenamefont{Takabayashi,
  Ganin, Jegli{\v{c}}, Ar{\v{c}}on, Takano, Iwasa, Ohishi, Takata, Takeshita,
  Prassides et~al.}}]{TakabayashiScience}
\bibinfo{author}{\bibfnamefont{Y.}~\bibnamefont{Takabayashi}},
  \bibinfo{author}{\bibfnamefont{A.~Y.} \bibnamefont{Ganin}},
  \bibinfo{author}{\bibfnamefont{P.}~\bibnamefont{Jegli{\v{c}}}},
  \bibinfo{author}{\bibfnamefont{D.}~\bibnamefont{Ar{\v{c}}on}},
  \bibinfo{author}{\bibfnamefont{T.}~\bibnamefont{Takano}},
  \bibinfo{author}{\bibfnamefont{Y.}~\bibnamefont{Iwasa}},
  \bibinfo{author}{\bibfnamefont{Y.}~\bibnamefont{Ohishi}},
  \bibinfo{author}{\bibfnamefont{M.}~\bibnamefont{Takata}},
  \bibinfo{author}{\bibfnamefont{N.}~\bibnamefont{Takeshita}},
  \bibinfo{author}{\bibfnamefont{K.}~\bibnamefont{Prassides}},
  \bibnamefont{et~al.}, \bibinfo{journal}{Science}
  \textbf{\bibinfo{volume}{323}}, \bibinfo{pages}{1585} (\bibinfo{year}{2009}).

\bibitem[{\citenamefont{Ganin et~al.}(2010)\citenamefont{Ganin, Takabayashi,
  Jegli{\v{c}}, Ar{\v{c}}on, Poto{\v{c}}nik, Baker, Ohishi, McDonald, Tzirakis,
  McLennan et~al.}}]{Ganin2010Nature}
\bibinfo{author}{\bibfnamefont{A.~Y.} \bibnamefont{Ganin}},
  \bibinfo{author}{\bibfnamefont{Y.}~\bibnamefont{Takabayashi}},
  \bibinfo{author}{\bibfnamefont{P.}~\bibnamefont{Jegli{\v{c}}}},
  \bibinfo{author}{\bibfnamefont{D.}~\bibnamefont{Ar{\v{c}}on}},
  \bibinfo{author}{\bibfnamefont{A.}~\bibnamefont{Poto{\v{c}}nik}},
  \bibinfo{author}{\bibfnamefont{P.~J.} \bibnamefont{Baker}},
  \bibinfo{author}{\bibfnamefont{Y.}~\bibnamefont{Ohishi}},
  \bibinfo{author}{\bibfnamefont{M.~T.} \bibnamefont{McDonald}},
  \bibinfo{author}{\bibfnamefont{M.~D.} \bibnamefont{Tzirakis}},
  \bibinfo{author}{\bibfnamefont{A.}~\bibnamefont{McLennan}},
  \bibnamefont{et~al.}, \bibinfo{journal}{Nature}
  \textbf{\bibinfo{volume}{466}}, \bibinfo{pages}{221} (\bibinfo{year}{2010}).

\bibitem[{\citenamefont{Zadik et~al.}(2015)\citenamefont{Zadik, Takabayashi,
  Klupp, Colman, Ganin, Poto{\v{c}}nik, Jegli{\v{c}}, Ar{\v{c}}on, Matus,
  Kamar{\'a}s et~al.}}]{Zadik2015Science}
\bibinfo{author}{\bibfnamefont{R.~H.} \bibnamefont{Zadik}},
  \bibinfo{author}{\bibfnamefont{Y.}~\bibnamefont{Takabayashi}},
  \bibinfo{author}{\bibfnamefont{G.}~\bibnamefont{Klupp}},
  \bibinfo{author}{\bibfnamefont{R.~H.} \bibnamefont{Colman}},
  \bibinfo{author}{\bibfnamefont{A.~Y.} \bibnamefont{Ganin}},
  \bibinfo{author}{\bibfnamefont{A.}~\bibnamefont{Poto{\v{c}}nik}},
  \bibinfo{author}{\bibfnamefont{P.}~\bibnamefont{Jegli{\v{c}}}},
  \bibinfo{author}{\bibfnamefont{D.}~\bibnamefont{Ar{\v{c}}on}},
  \bibinfo{author}{\bibfnamefont{P.}~\bibnamefont{Matus}},
  \bibinfo{author}{\bibfnamefont{K.}~\bibnamefont{Kamar{\'a}s}},
  \bibnamefont{et~al.}, \bibinfo{journal}{Sci. Adv.}
  \textbf{\bibinfo{volume}{1}}, \bibinfo{pages}{e1500059}
  (\bibinfo{year}{2015}).

\bibitem[{\citenamefont{Mitrano et~al.}(2016)\citenamefont{Mitrano, Cantaluppi,
  Nicoletti, Kaiser, Perucchi, Lupi, Di~Pietro, Pontiroli, Ricc{\`o}, Clark
  et~al.}}]{Mitrano_Nature2016}
\bibinfo{author}{\bibfnamefont{M.}~\bibnamefont{Mitrano}},
  \bibinfo{author}{\bibfnamefont{A.}~\bibnamefont{Cantaluppi}},
  \bibinfo{author}{\bibfnamefont{D.}~\bibnamefont{Nicoletti}},
  \bibinfo{author}{\bibfnamefont{S.}~\bibnamefont{Kaiser}},
  \bibinfo{author}{\bibfnamefont{A.}~\bibnamefont{Perucchi}},
  \bibinfo{author}{\bibfnamefont{S.}~\bibnamefont{Lupi}},
  \bibinfo{author}{\bibfnamefont{P.}~\bibnamefont{Di~Pietro}},
  \bibinfo{author}{\bibfnamefont{D.}~\bibnamefont{Pontiroli}},
  \bibinfo{author}{\bibfnamefont{M.}~\bibnamefont{Ricc{\`o}}},
  \bibinfo{author}{\bibfnamefont{S.~R.} \bibnamefont{Clark}},
  \bibnamefont{et~al.}, \bibinfo{journal}{Nature}
  \textbf{\bibinfo{volume}{530}}, \bibinfo{pages}{461} (\bibinfo{year}{2016}).

\bibitem[{\citenamefont{Capone et~al.}(2002)\citenamefont{Capone, Fabrizio,
  Castellani, and Tosatti}}]{CaponeScience}
\bibinfo{author}{\bibfnamefont{M.}~\bibnamefont{Capone}},
  \bibinfo{author}{\bibfnamefont{M.}~\bibnamefont{Fabrizio}},
  \bibinfo{author}{\bibfnamefont{C.}~\bibnamefont{Castellani}},
  \bibnamefont{and} \bibinfo{author}{\bibfnamefont{E.}~\bibnamefont{Tosatti}},
  \bibinfo{journal}{Science} \textbf{\bibinfo{volume}{296}},
  \bibinfo{pages}{2364} (\bibinfo{year}{2002}).

\bibitem[{\citenamefont{Capone et~al.}(2009)\citenamefont{Capone, Fabrizio,
  Castellani, and Tosatti}}]{Capone_RMP2009}
\bibinfo{author}{\bibfnamefont{M.}~\bibnamefont{Capone}},
  \bibinfo{author}{\bibfnamefont{M.}~\bibnamefont{Fabrizio}},
  \bibinfo{author}{\bibfnamefont{C.}~\bibnamefont{Castellani}},
  \bibnamefont{and} \bibinfo{author}{\bibfnamefont{E.}~\bibnamefont{Tosatti}},
  \bibinfo{journal}{Rev. Mod. Phys.} \textbf{\bibinfo{volume}{81}},
  \bibinfo{pages}{943} (\bibinfo{year}{2009}).

\bibitem[{\citenamefont{Nomura et~al.}(2015)\citenamefont{Nomura, Sakai,
  Capone, and Arita}}]{NomuraScience}
\bibinfo{author}{\bibfnamefont{Y.}~\bibnamefont{Nomura}},
  \bibinfo{author}{\bibfnamefont{S.}~\bibnamefont{Sakai}},
  \bibinfo{author}{\bibfnamefont{M.}~\bibnamefont{Capone}}, \bibnamefont{and}
  \bibinfo{author}{\bibfnamefont{R.}~\bibnamefont{Arita}},
  \bibinfo{journal}{Sci. Adv.} \textbf{\bibinfo{volume}{1}},
  \bibinfo{pages}{e1500568} (\bibinfo{year}{2015}).

\bibitem[{\citenamefont{Tahara and Imada}(2008)}]{TaharaVMC}
\bibinfo{author}{\bibfnamefont{D.}~\bibnamefont{Tahara}} \bibnamefont{and}
  \bibinfo{author}{\bibfnamefont{M.}~\bibnamefont{Imada}}, \bibinfo{journal}{J.
  Phys. Soc. Jpn.} \textbf{\bibinfo{volume}{77}}, \bibinfo{pages}{114701}
  (\bibinfo{year}{2008}).

\bibitem[{\citenamefont{Misawa and Imada}(2014{\natexlab{b}})}]{misawaHubbard}
\bibinfo{author}{\bibfnamefont{T.}~\bibnamefont{Misawa}} \bibnamefont{and}
  \bibinfo{author}{\bibfnamefont{M.}~\bibnamefont{Imada}},
  \bibinfo{journal}{Phys. Rev. B} \textbf{\bibinfo{volume}{90}},
  \bibinfo{pages}{115137} (\bibinfo{year}{2014}{\natexlab{b}}).

\bibitem[{mVM()}]{mVMC}
\bibinfo{note}{~http://ma.cms-initiative.jp/en/application-list/mvmc}.

\bibitem[{SM()}]{SM}
\bibinfo{note}{See Supplemental Materials for details of $J_{\rm pair}$ and
  $J_{\rm ex}$ dependence of the superconducting correlations, doping
  dependence of $D_{\nu}$, size dependence of the spin structure factors,
  short-range antiferromagnetic correlations of disordered orbital, low-energy
  excited state, and spin and charge structure factors in the two-orbital
  model.}

\bibitem[{\citenamefont{Ohgoe and Imada}()}]{Ohgoe_2017}
\bibinfo{author}{\bibfnamefont{T.}~\bibnamefont{Ohgoe}} \bibnamefont{and}
  \bibinfo{author}{\bibfnamefont{M.}~\bibnamefont{Imada}},
  \bibinfo{howpublished}{arXiv:1703.08899}.

\bibitem[{\citenamefont{Nomura et~al.}(2012)\citenamefont{Nomura, Nakamura, and
  Arita}}]{NomuraPRB}
\bibinfo{author}{\bibfnamefont{Y.}~\bibnamefont{Nomura}},
  \bibinfo{author}{\bibfnamefont{K.}~\bibnamefont{Nakamura}}, \bibnamefont{and}
  \bibinfo{author}{\bibfnamefont{R.}~\bibnamefont{Arita}},
  \bibinfo{journal}{Phys. Rev. B} \textbf{\bibinfo{volume}{85}},
  \bibinfo{pages}{155452} (\bibinfo{year}{2012}).

\bibitem[{\citenamefont{Gutzwiller}(1963)}]{Gutzwiller}
\bibinfo{author}{\bibfnamefont{M.~C.} \bibnamefont{Gutzwiller}},
  \bibinfo{journal}{Phys. Rev. Lett.} \textbf{\bibinfo{volume}{10}},
  \bibinfo{pages}{159} (\bibinfo{year}{1963}).

\bibitem[{\citenamefont{Jastrow}(1955)}]{Jastrow}
\bibinfo{author}{\bibfnamefont{R.}~\bibnamefont{Jastrow}},
  \bibinfo{journal}{Phys. Rev.} \textbf{\bibinfo{volume}{98}},
  \bibinfo{pages}{1479} (\bibinfo{year}{1955}).

\bibitem[{\citenamefont{Capello et~al.}(2005)\citenamefont{Capello, Becca,
  Fabrizio, Sorella, and Tosatti}}]{CapelloJastrow}
\bibinfo{author}{\bibfnamefont{M.}~\bibnamefont{Capello}},
  \bibinfo{author}{\bibfnamefont{F.}~\bibnamefont{Becca}},
  \bibinfo{author}{\bibfnamefont{M.}~\bibnamefont{Fabrizio}},
  \bibinfo{author}{\bibfnamefont{S.}~\bibnamefont{Sorella}}, \bibnamefont{and}
  \bibinfo{author}{\bibfnamefont{E.}~\bibnamefont{Tosatti}},
  \bibinfo{journal}{Phys. Rev. Lett.} \textbf{\bibinfo{volume}{94}},
  \bibinfo{pages}{026406} (\bibinfo{year}{2005}).

\bibitem[{\citenamefont{Gros}(1989)}]{gros1989physics}
\bibinfo{author}{\bibfnamefont{C.}~\bibnamefont{Gros}}, \bibinfo{journal}{Ann.
  Phys.} \textbf{\bibinfo{volume}{189}}, \bibinfo{pages}{53}
  (\bibinfo{year}{1989}).

\bibitem[{\citenamefont{Bajdich et~al.}(2008)\citenamefont{Bajdich, Mitas,
  Wagner, and Schmidt}}]{BajdichPRB}
\bibinfo{author}{\bibfnamefont{M.}~\bibnamefont{Bajdich}},
  \bibinfo{author}{\bibfnamefont{L.}~\bibnamefont{Mitas}},
  \bibinfo{author}{\bibfnamefont{L.~K.} \bibnamefont{Wagner}},
  \bibnamefont{and} \bibinfo{author}{\bibfnamefont{K.~E.}
  \bibnamefont{Schmidt}}, \bibinfo{journal}{Phys. Rev. B}
  \textbf{\bibinfo{volume}{77}}, \bibinfo{pages}{115112}
  (\bibinfo{year}{2008}).

\bibitem[{\citenamefont{Sorella}(2001)}]{Sorella_PRB2001}
\bibinfo{author}{\bibfnamefont{S.}~\bibnamefont{Sorella}},
  \bibinfo{journal}{Phys. Rev. B} \textbf{\bibinfo{volume}{64}},
  \bibinfo{pages}{024512} (\bibinfo{year}{2001}).

\bibitem[{\citenamefont{Suhl et~al.}(1959)\citenamefont{Suhl, Matthias, and
  Walker}}]{Suhl_PRL1959}
\bibinfo{author}{\bibfnamefont{H.}~\bibnamefont{Suhl}},
  \bibinfo{author}{\bibfnamefont{B.~T.} \bibnamefont{Matthias}},
  \bibnamefont{and} \bibinfo{author}{\bibfnamefont{L.~R.}
  \bibnamefont{Walker}}, \bibinfo{journal}{Phys. Rev. Lett.}
  \textbf{\bibinfo{volume}{3}}, \bibinfo{pages}{552} (\bibinfo{year}{1959}).

\bibitem[{\citenamefont{Kondo}(1963)}]{Kondo_PTP1963}
\bibinfo{author}{\bibfnamefont{J.}~\bibnamefont{Kondo}},
  \bibinfo{journal}{Prog. Theor. Phys.} \textbf{\bibinfo{volume}{29}},
  \bibinfo{pages}{1} (\bibinfo{year}{1963}).

\bibitem[{\citenamefont{Kasahara et~al.}(2014)\citenamefont{Kasahara, Takeuchi,
  Itou, Zadik, Takabayashi, Ganin, Ar\ifmmode~\check{c}\else \v{c}\fi{}on,
  Rosseinsky, Prassides, and Iwasa}}]{KasaharaPRB2014}
\bibinfo{author}{\bibfnamefont{Y.}~\bibnamefont{Kasahara}},
  \bibinfo{author}{\bibfnamefont{Y.}~\bibnamefont{Takeuchi}},
  \bibinfo{author}{\bibfnamefont{T.}~\bibnamefont{Itou}},
  \bibinfo{author}{\bibfnamefont{R.~H.} \bibnamefont{Zadik}},
  \bibinfo{author}{\bibfnamefont{Y.}~\bibnamefont{Takabayashi}},
  \bibinfo{author}{\bibfnamefont{A.~Y.} \bibnamefont{Ganin}},
  \bibinfo{author}{\bibfnamefont{D.}~\bibnamefont{Ar\ifmmode~\check{c}\else
  \v{c}\fi{}on}}, \bibinfo{author}{\bibfnamefont{M.~J.}
  \bibnamefont{Rosseinsky}},
  \bibinfo{author}{\bibfnamefont{K.}~\bibnamefont{Prassides}},
  \bibnamefont{and} \bibinfo{author}{\bibfnamefont{Y.}~\bibnamefont{Iwasa}},
  \bibinfo{journal}{Phys. Rev. B} \textbf{\bibinfo{volume}{90}},
  \bibinfo{pages}{014413} (\bibinfo{year}{2014}).

\bibitem[{\citenamefont{Kitaev}(2006)}]{Kitaev}
\bibinfo{author}{\bibfnamefont{A.}~\bibnamefont{Kitaev}},
  \bibinfo{journal}{Ann. Phys.} \textbf{\bibinfo{volume}{321}},
  \bibinfo{pages}{2} (\bibinfo{year}{2006}).

\bibitem[{\citenamefont{Hoshino and Werner}(2017)}]{HoshinoPRL}
\bibinfo{author}{\bibfnamefont{S.}~\bibnamefont{Hoshino}} \bibnamefont{and}
  \bibinfo{author}{\bibfnamefont{P.}~\bibnamefont{Werner}},
  \bibinfo{journal}{Phys. Rev. Lett.} \textbf{\bibinfo{volume}{118}},
  \bibinfo{pages}{177002} (\bibinfo{year}{2017}).

\bibitem[{\citenamefont{Koga and Werner}(2015)}]{KogaPRB}
\bibinfo{author}{\bibfnamefont{A.}~\bibnamefont{Koga}} \bibnamefont{and}
  \bibinfo{author}{\bibfnamefont{P.}~\bibnamefont{Werner}},
  \bibinfo{journal}{Physical Review B} \textbf{\bibinfo{volume}{91}},
  \bibinfo{pages}{085108} (\bibinfo{year}{2015}).

\end{thebibliography}

\end{document}